# Decoupling Reconnaissance and Exploitation: Measuring the Capability Boundaries of LLM-Based Web Penetration Testing

Liwei Yu[1[0009-0000-5272-2468]], Shuo Li[1[0009-0004-3417-1147]], Ming Zhou[3[0009-0005-6873-5710]], Ge Chu[4[0009-0005-1172-050X]] and Yan Guo[1,2] ✉ [0000-0003-2091-7732]

[1] University of Science and Technology of China, Hefei, China
[2] Suzhou Institute for Advanced Research, University of Science and Technology of China
[3] Nanjing University of Science and Technology, Nanjing, China
[4] Research Institute, Runjian Co., Ltd, China
guoyan@ustc.edu.cn

**Abstract.** Large Language Models (LLMs) have shown promise for automated penetration testing, yet existing end-to-end black-box evaluations are highly susceptible to error cascading: failures in early reconnaissance can mask an agent's actual ability to exploit vulnerabilities. To more accurately characterize these capabilities, we propose a two-stage decoupled evaluation framework that separates exploit execution from reconnaissance. Using ground-truth injection and knowledge-driven ablation across 70 high-fidelity web vulnerability testbeds, our framework isolates exploitation performance from reconnaissance noise. We empirically evaluate five open-source penetration-testing agents, covering multi-agent, monolithic, and graph-driven architectures, on a strictly aligned subset of 50 representative vulnerabilities. The results reveal a substantial capability gap. With accurate vulnerability context, agents achieve a functional success rate of up to 90.0%, whereas autonomous reconnaissance, measured by targeted vulnerability recall, plateaus at approximately 50.0%, primarily due to failures in parsing unstructured telemetry. Cross-architectural analysis further reveals distinct capability niches: multi-agent isolation is more effective for long-sequence interactions such as de-serialization, while monolithic and graph-driven designs perform better on short-chain injections and cross-session access-control vulnerabilities, respectively. This decoupled evaluation work provides a fine-grained benchmarking protocol and an empirical basis for designing next-generation automated offensive security agents.

**Keywords:** Automated Penetration Testing, Web Vulnerability, Large Language Models, Decoupled Evaluation, Ground-Truth Injection

## 1 Introduction

Recent advances in Large Language Models (LLMs) for code generation and logical reasoning have accelerated the development of autonomous cybersecurity systems. Prior work, such as PentestGPT [1] has shown the potential of LLMs for interpreting security logs, while Fang et al. [3] demonstrated that agents can exploit 1-day vulnerabilities in constrained settings. Frameworks including AutoPentest [6] and AutoAttacker [18] further integrate external tools to support asset mapping and single-step exploitation. In parallel, recent studies on network threat characterization and intrusion detection have emphasized the complexity of real-world cyber environments: Zhou et al. [19] characterized network threats against industrial control systems using honeypot technology, while SecureNet-AWMI improves network intrusion detection through optimal feature selection under noisy and imbalanced traffic conditions [20]. These efforts highlight that practical security scenarios involve unstable observations and adversarial traffic patterns. However, existing LLM-based penetration-testing agents still exhibit substantially degraded effectiveness in realistic environments involving WAF interception, non-standard error messages, and network instability [16]. As a result, the community still lacks a precise understanding of agent capability boundaries in adversarial settings, as well as a reliable way to attribute failures to base-model limitations or control-flow design flaws.

This evaluation bottleneck stems largely from the limitations of end-to-end black-box testing [5]. In such settings, penetration-testing stages are tightly coupled, making evaluations vulnerable to error cascading: a failure in early reconnaissance can directly prevent subsequent exploitation. This masking effect causes agents' exploitation capabilities to be underestimated due to preceding perception errors, producing statistical false negatives that impede the architectural improvement of automated penetration-testing frameworks.

To address these limitations, we propose a two-stage decoupled evaluation framework that separates autonomous reconnaissance from vulnerability exploitation. When reconnaissance fails, a ground-truth injection mechanism supplies accurate vulnerability context, breaking the error-cascading chain and creating a controlled setting that isolates exploitation performance from reconnaissance noise. Using 70 high-fidelity web vulnerability targets, we benchmark five open-source penetration-testing frameworks with heterogeneous architectures and provide fine-grained measurements of their performance across diverse vulnerability classes.

This work makes four contributions. First, we introduce a two-stage decoupled evaluation framework. By incorporating ground-truth injection, the framework mitigates the evaluation bias caused by error cascading in traditional end-to-end black-box testing and establishes a fine-grained quantitative benchmarking protocol for LLM-based penetration-testing agents. Second, we construct a high-fidelity benchmark dataset. To address the lack of standardized testbeds, we curate and reconstruct 70 representative web vulnerability environments from Vulhub and Vulfocus, together with standardized prior-knowledge contexts that emphasize adversarial complexity and reproducibility. Third, we quantify the capability boundaries of current penetration-testing agents. Large-scale empirical results show that agents can achieve strong exploitation performance when given accurate prior knowledge, with a Functional Success Rate (FSR) of up to 90.0%, but remain bottlenecked by unstructured environmental perception, with Targeted Vulnerability Recall (TVR) generally around 50%. Fourth, we provide systematic architectural attribution and design implications. Through reverse analysis of agent control flows, we identify key architectural flaws associated with failures and derive recommendations for next-generation composite architectures incorporating physical state isolation.

The remainder of this paper is organized as follows. Section 2 reviews related work. Section 3 presents the design and implementation of the two-stage decoupled evaluation framework. Section 4 describes the benchmark construction and reports quantitative results for the five agents. Section 5 analyzes architectural trade-offs across different frameworks. Section 6 concludes the paper.

## 2 Related Work

**LLMs in Offensive Security.** Early studies on offensive applications of Large Language Models (LLMs) established the feasibility of automating penetration-testing tasks. PentestGPT [1] showed that LLMs can parse unstructured security logs and support security reasoning. Subsequent work [2, 3] demonstrated that agents can autonomously exploit 1-day vulnerabilities in sandboxed environments, while frameworks such as AutoPentest [6], AutoAttacker [18], and Incalmo [13] extended automated exploitation to assumed-breach scenarios and multi-host red teaming. Across these studies, however, a consistent limitation emerges: without explicit vulnerability context, exploit generation success rates decline sharply, highlighting the importance of external knowledge injection.

**Agent Architectures and Tool Augmentation.** In complex penetration scenarios, agents must process noisy target telemetry without losing task intent or degrading reasoning quality [15]. Two architectural paradigms have been explored to address this challenge. Multi-agent systems, such as CurriculumPT [17] and MindAgent [4], use role decoupling to manage complex interactions and often incorporate verbal reflection for iterative self-correction [12]. Neuro-symbolic approaches instead rely on structured attack trees to enforce execution constraints and improve determinism [8, 14]. In parallel, dynamic knowledge injection and standardized tool interfaces, such as CRAKEN [11] and ToolLLM [10], have been introduced to reduce low-level execution errors, including memory-offset mistakes and tool-invocation hallucinations, thereby supporting more reliable attack planning [7].

**Limitations of End-to-End Evaluation.** Although benchmarks such as AgentBench [9] and WebArena [21] provide dynamic environments, security-oriented evaluation remains limited by the prevailing end-to-end black-box paradigm. Penetration testing involves sequential stages, and tightly coupling reconnaissance with exploitation creates a strong masking effect: failures in early fingerprinting can trigger error cascading, causing downstream exploitation capability to be recorded as failure [5]. This makes it difficult to distinguish factual knowledge deficits from reasoning or control-flow flaws. These limitations motivate a decoupled evaluation protocol that can independently assess reconnaissance and exploitation capabilities.

# 3 Evaluation Methodology

To characterize the capability boundaries of LLM-based penetration-testing agents and identify bottlenecks across control-flow architectures, we propose a two-stage decoupled evaluation framework with ground-truth injection. By separating stage-wise errors, the framework mitigates the error cascading and masking effects inherent in end-to-end (E2E) evaluation, enabling fine-grained measurement of exploitation capability independent of reconnaissance noise.

## 3.1 Decoupled Framework Architecture

Manual penetration testing typically proceeds through information gathering, vulnerability scanning, proof-of-concept verification, and exploitation. To avoid error amplification across this sequential pipeline, we aggregate these steps into two orthogonal evaluation phases. **Fig. 1** shows the architecture comparison between the end-to-end evaluation and our framework.

**Reconnaissance Phase.** This phase covers information gathering and vulnerability scanning. It evaluates an agent's ability to invoke probing tools, interpret unstructured telemetry, and identify the target vulnerability, such as by producing the correct CVE identifier.

**Exploitation Phase.** This phase covers vulnerability verification and exploitation. Starting from an explicit target context, the agent is evaluated on payload construction, dynamic error correction, and exploit execution within the target environment.

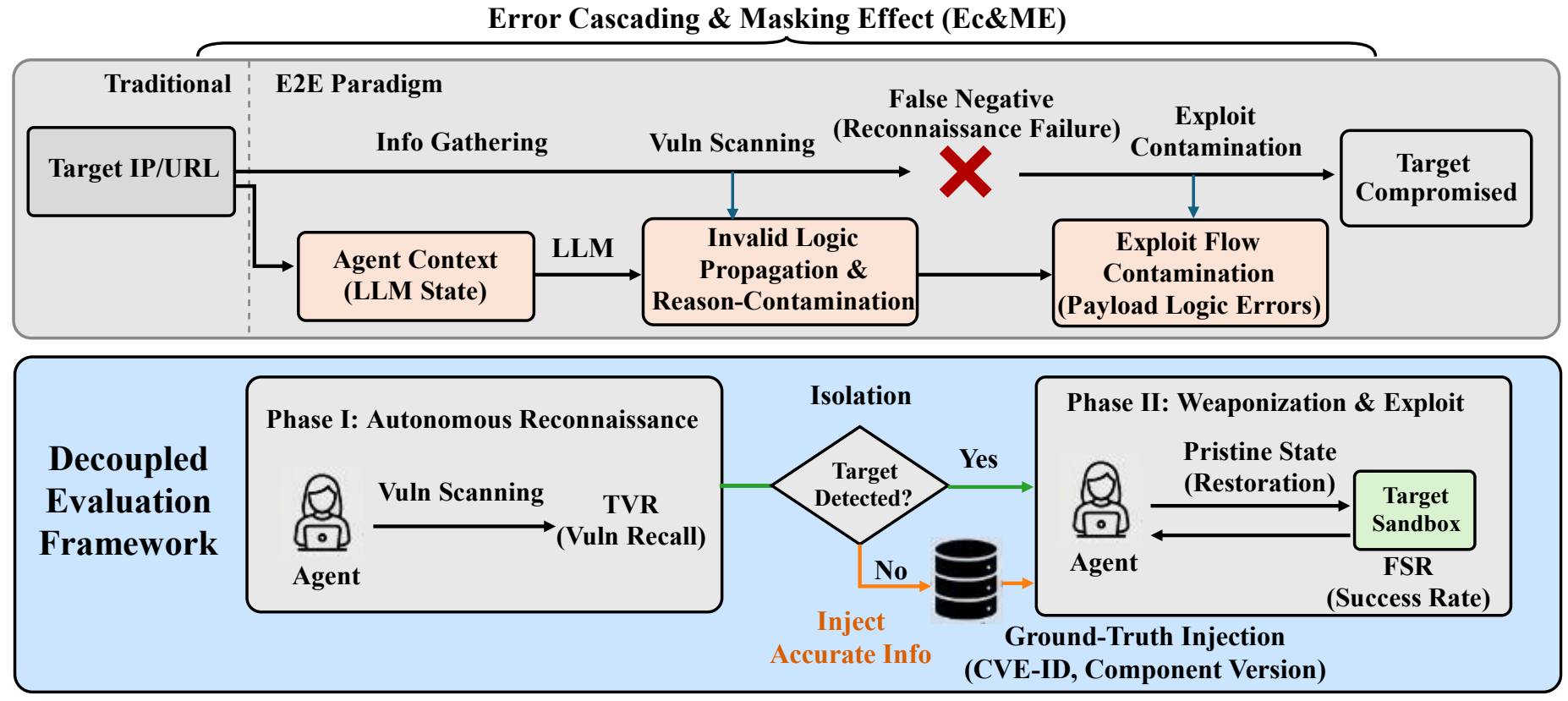

**Fig. 1.** Architectural comparison between the conventional end-to-end evaluation and the proposed two-stage decoupled framework.

## 3.2 Core Evaluation Mechanisms

Ground-truth injection is used to break error propagation between reconnaissance and exploitation. If an agent fails to identify the correct vulnerability during reconnaissance, the framework pauses the current execution flow and injects the authentic vulnerability identifier into the prompt for the subsequent exploitation phase. This controlled design ensures that the agent enters exploitation with the required prerequisite knowledge, thereby isolating execution capability from reconnaissance errors. The mechanism introduces minimal overhead, as it only requires prompt reconstruction and no additional testbed preparation.

To analyze the causes of tool-invocation and code-generation failures, we introduce a Verified Empirical Knowledge Base (VEKB) and evaluate agents under two settings: zero-knowledge and knowledge-augmented. **Fig. 2** illustrates the micro-architecture of this dual-track ablation engine. The former relies solely on the model's pre-trained parameters, whereas the latter provides standardized Proof-of-Concept (PoC) templates. Comparing these settings allows us to isolate factual knowledge deficits and examine architectural differences in knowledge utilization and generalization.

Beyond aggregate success rates, we categorize testbeds by vulnerability type according to CWE, including de-serialization, injection, and broken access control. We then construct a cross-mapping matrix between agent architectures and vulnerability classes, enabling quantitative analysis of how different control-flow designs perform across specific flaw types. This analysis provides empirical support for selecting and designing agent architectures under realistic penetration-testing scenarios.

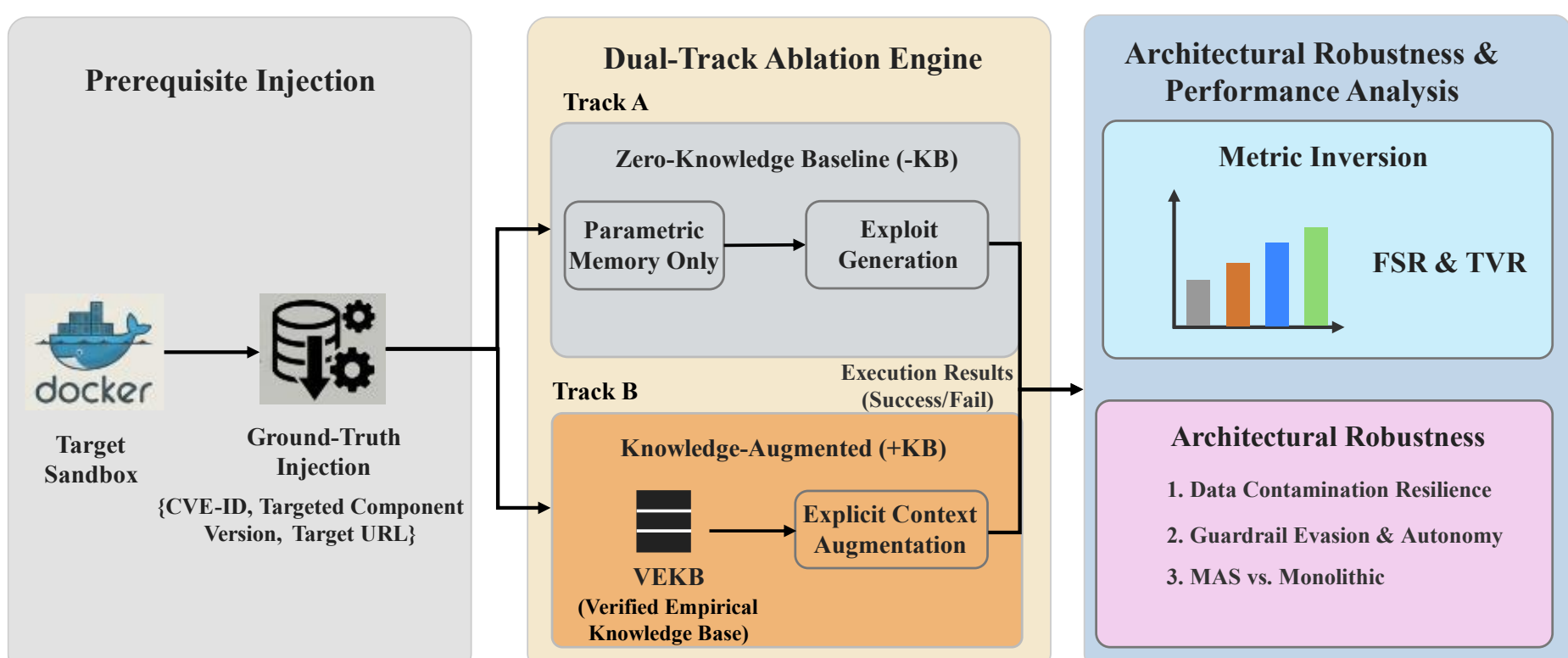

**Fig. 2.** Micro-architecture of the decoupled evaluation engine, illustrating knowledge-guided ablation and phase-wise capability attribution.

# 4 Experiments

## 4.1 Experimental Setup

To evaluate code generation and environmental adaptability across varying logical depths, we construct a high-fidelity dataset of 70 representative web vulnerabilities based on the OWASP Top 10 and CWE standards. As shown in **Table 1**, the dataset covers injection flaws, configuration errors, and logic-related vulnerabilities, enabling systematic assessment of payload construction under dynamic filtering, long-sequence interaction, and zero-shot generalization.

**Table 1.** Vulnerability taxonomy and class distribution in the evaluation dataset.

| CWE | Count | Ratio | Typical Examples |
|---|---|---|---|
| CWE-74 | 27 | 38.6% | SQL Injection, OS Command Injection, SSTI |
| CWE-284 | 13 | 18.6% | API Insecure Direct Object Reference (IDOR), Redis/Docker Unauthorized Access |
| CWE-22 | 13 | 18.6% | Arbitrary File Read, Insecure File Upload |
| CWE-502 | 12 | 17.1% | Weblogic Deserialization, Fastjson RCE |
| CWE-16 | 5 | 7.1% | Business Logic Bypass, Sensitive Information Disclosure |

All experiments are conducted in a physically isolated virtual sandbox. The attack node runs Kali Linux, hosts the evaluated penetration-testing frameworks, and orchestrates native security tools. The victim node contains 70 independently deployed Docker containers. To prevent cross-run state contamination, such as residual WebShells from long-sequence interactions, the orchestration engine destroys and rebuilds the corresponding container before each independent evaluation, ensuring a pristine initial state.

We adopt a tiered evaluation protocol consisting of core ablation studies and generalization verification. In Phase I, autonomous reconnaissance requires the agent to independently output the target CVE identifier. If reconnaissance fails, Phase II triggers ground-truth injection by providing the correct vulnerability context, thereby isolating exploitation performance from preceding reconnaissance errors.

We select five representative open-source penetration-testing frameworks with heterogeneous architectures as baselines; their architectural characteristics are summarized in **Table 2**. We characterize each framework along four dimensions: *Architecture*, which captures the high-level design paradigm, including monolithic and multi-agent designs; *Reasoning Mode*, which describes the mechanism for task planning and execution; *State Management*, which specifies how context is preserved and isolated during long-horizon engagements; and *Knowledge/Tool Interface*, which defines how external tools and threat intelligence are integrated.

For the core ablation study, Hexstrike-AI and CyberStrikeAI are evaluated on all 70 testbeds while controlling for foundation model scale, namely GPT-5 and Gemini 3 Pro, and prior-knowledge-base settings. For generalization verification, all five frameworks are evaluated with Gemini 3 Pro on a strictly aligned subset of 50 representative vulnerabilities.

**Table 2.** Architectural characteristics of the five automated penetration-testing agents.

| Framework | Architecture | Reasoning Mode | State Management | Knowledge/Tool Interface |
|---|---|---|---|---|
| Hexstrike-AI | MCP-driven MAS | Tool-Augmented | Isolated MCP | MCP Server |
| CyberStrikeAI | Monolithic | Task Delegation | Shared Space | OWASP Repositories |
| RedAmon | Graph-driven MAS | Fan-out/Fan-in | Knowledge Graph | PTY & OSINT |
| NeuroSploit | Constrained Agent | Symbolic Tree | Containerized | Prompt Guards |
| Decepticon | Graph-driven MAS | Collab. Workflow | Global Topology | Auto-Toolchain |

Following the two-stage decoupled design, we define two orthogonal metrics. The first is Targeted Vulnerability Recall (TVR), which measures reconnaissance capability. It is defined as the ratio of correctly identified target vulnerabilities ($N_{detected}$) to the total number of testbeds ($N_{total}$).

$$TVR = \frac{N_{detected}}{N_{total}} \tag{1}$$

The second is Functional Success Rate (FSR), which measures exploitation capability after ground-truth injection. It is defined as the ratio of instances that successfully trigger the expected security impact, such as establishing a reverse shell, ($N_{exploit}$) to the total number of instances entering the reproduction phase ($N_{reproduce}$).

$$FSR = \frac{N_{exploit}}{N_{reproduce}} \tag{2}$$

### 4.2 Results and Analysis

In this section, we use TVR and FSR to quantify performance boundaries in the 70-vulnerability ablation study and the 50-vulnerability generalization verification.

**Ablation of Core Architectures and Knowledge States.** To analyze the effects of foundation models, control-flow architectures, and external knowledge, we evaluate Hexstrike-AI and CyberStrikeAI on the 70 testbeds. The results are shown in **Table 3**, while **Fig. 3** visualizes the comparative trends of TVR and FSR across different configurations.

**Table 3.** Performance ablation of core penetration-testing agents on the full evaluation dataset.

| Framework | Base Model | With VEKB (TVR) | With VEKB (FSR) | Without KB (TVR) | Without KB (FSR) |
|---|---|---|---|---|---|
| Hexstrike-AI | Gemini 3 Pro | 61(87.1%) | 57(81.4%) | 38(54.3%) | 43(61.4%) |
| Hexstrike-AI | GPT-5 High | 56(80.0%) | 54(77.1%) | 33(47.1%) | 38(54.3%) |
| CyberstrikeAI | Gemini 3 Pro | 59(84.3%) | 54(77.1%) | 43(61.4%) | 39(55.7%) |
| CyberstrikeAI | GPT-5 High | 61(87.1%) | 45(64.3%) | 32(45.7%) | 31(44.3%) |

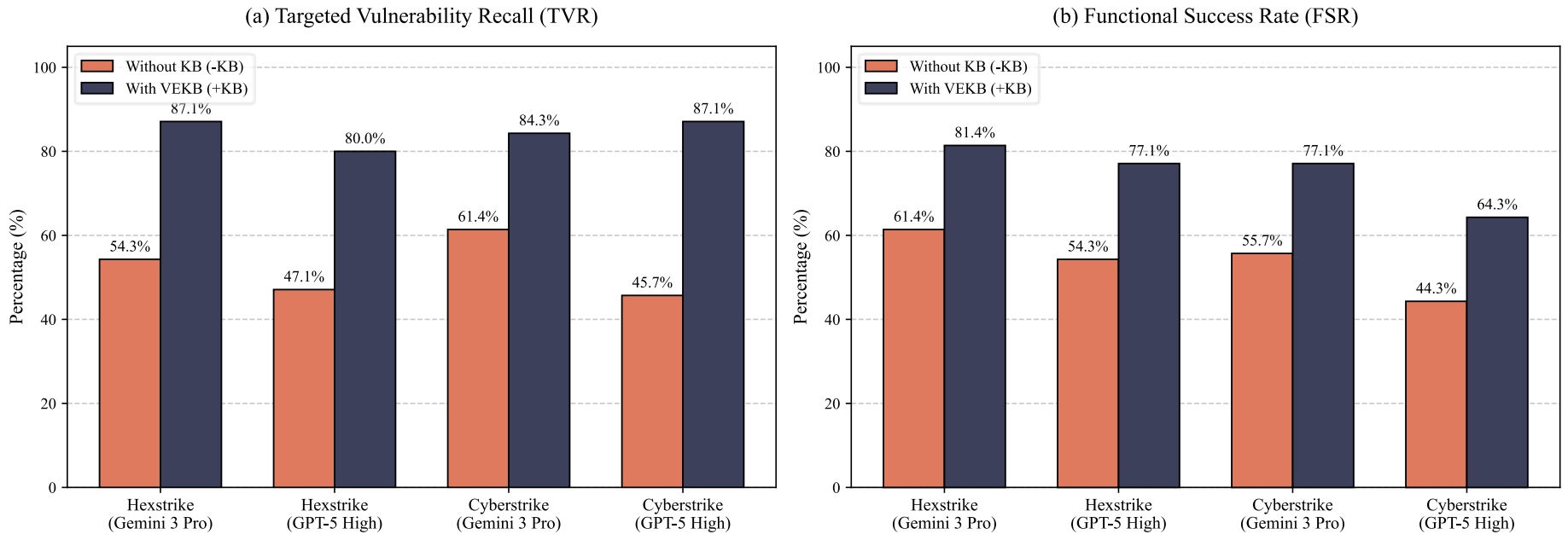

**Fig. 3.** Comparison of targeted vulnerability recall (TVR) and functional success rate (FSR) across different variable configurations.

The Verified Empirical Knowledge Base (VEKB) is a critical factor in overcoming code-execution bottlenecks. Without the knowledge base, FSR for both architectures ranges from 44.3% to 61.4%, with frequent failures in payload closure and memory-offset calculation. After VEKB integration, performance improves consistently. For example, Hexstrike-AI with Gemini 3 Pro increases its FSR from 61.4% to 81.4%. This indicates that injecting verified empirical knowledge mitigates prior-knowledge deficits in low-level security-tool protocols.

The results expose a clear gap between environmental perception and exploit execution. Without knowledge-base assistance, Hexstrike-AI with Gemini 3 Pro achieves only 54.3% TVR, while its FSR reaches 61.4%. This inversion illustrates a key flaw of E2E evaluation: agents often miss vulnerabilities because they fail to extract relevant features from unstructured scanning logs, yet such failures are recorded as exploitation failures. After decoupling and ground-truth injection, agents demonstrate reproduction capabilities that exceed their autonomous reconnaissance performance.

We compare GPT-5 and Gemini 3 Pro under identical architectures. Larger or more capable base models do not necessarily yield better offensive performance. In the monolithic CyberStrikeAI framework, replacing Gemini 3 Pro with GPT-5 decreases FSR from 77.1% to 64.3%. Log analysis suggests that GPT-5 is more susceptible to attention dispersion when processing verbose error telemetry without state isolation. In contrast, the decoupled multi-agent architecture remains robust across both models. This suggests that architectural state isolation is a prerequisite for effectively scaling LLMs in offensive security tasks.

**Cross-Framework Generalization and Robustness.** To examine the generality of explicit knowledge augmentation and multi-agent design, we evaluate five heterogeneous frameworks using Gemini 3 Pro on the aligned subset of 50 high-complexity testbeds.

**Table 4.** Cross-framework generalization performance on 50 representative CVEs.

| Framework | With VEKB (TVR) | With VEKB (FSR) | Without KB (TVR) | Without KB (FSR) |
|---|---|---|---|---|
| Hexstrike-AI | 45(90.0%) | 45(90.0%) | 27(54.0%) | 33(66.0%) |
| CyberstrikeAI | 43(86.0%) | 40(80.0%) | 29(58.0%) | 29(58.0%) |
| RedAmon | 43(86.0%) | 37(74.0%) | 28(56.0%) | 29(58.0%) |
| Decepticon | 44(88.0%) | 35(70.0%) | 24(48.0%) | 24(48.0%) |
| NeuroSploit | 42(84.0%) | 31(62.0%) | 26(52.0%) | 27(54.0%) |

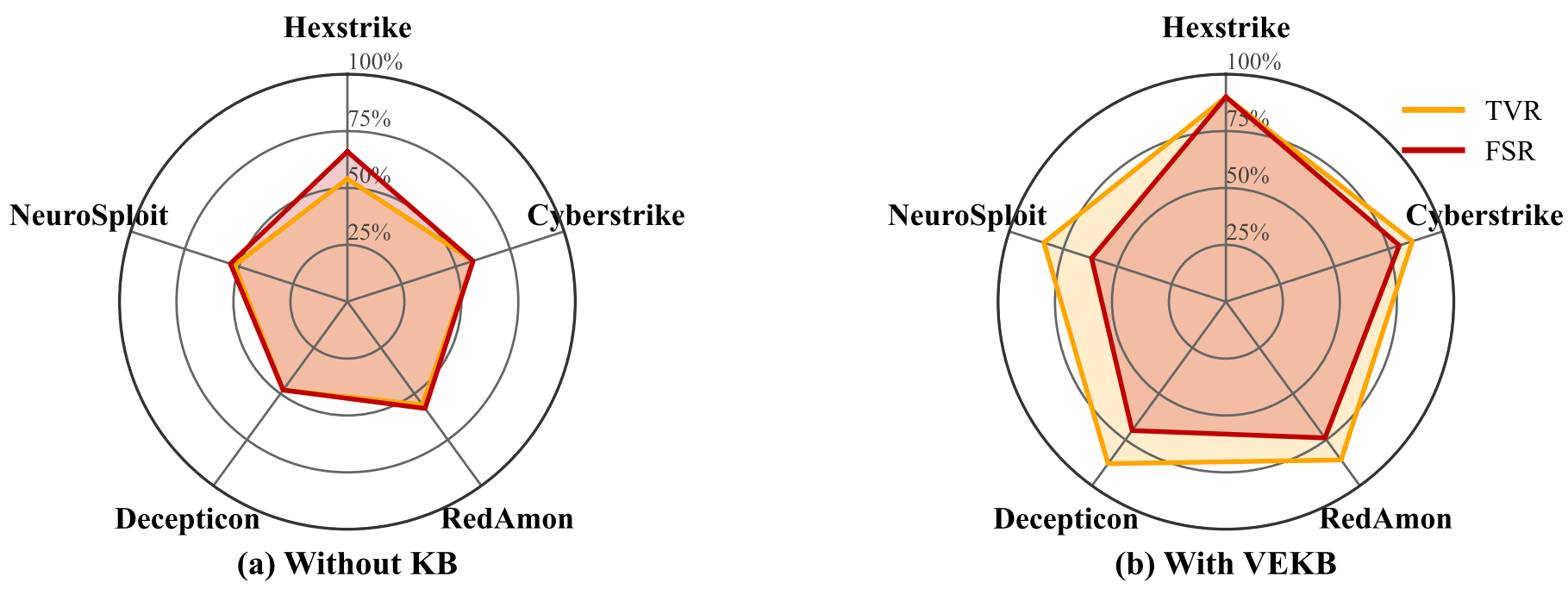

**Fig. 4.** Multidimensional comparison of generalization efficacy across five heterogeneous pentesting frameworks on the strictly aligned subset.

As shown in **Table 4**, VEKB integration substantially improves TVR across all five frameworks, all exceeding 84% and Hexstrike reaching 90.0%. This shows that matching standard fingerprints and version metadata during reconnaissance reduces omission rates regardless of the underlying tool-orchestration logic. Explicit knowledge injection therefore provides a general mechanism to reduce vulnerability-detection blind spots.

Although reconnaissance performance becomes more homogeneous after knowledge augmentation, FSR differs sharply across architectures. **Fig. 4** clearly visualizes this contrast across multiple dimensions. Hexstrike-AI, which uses the Model Context Protocol (MCP) for isolation, achieves the highest FSR of 90.0%. By contrast, NeuroSploit and Decepticon achieve 62.0% and 70.0%, respectively. Under the zero-knowledge baseline, we observe 51 masking instances in which agents fail reconnaissance but successfully exploit the target after ground-truth injection. This confirms that realistic vulnerability reproduction involves frequent network exceptions and noisy error telemetry. Multi-agent architectures with state isolation better prevent error logs from contaminating the primary planner, yielding stronger robustness than monolithic or graph-based designs in high-noise adversarial environments.

## 5 Discussion

**Capability boundaries of LLMs in web pentesting.** The decoupled evaluation reveals clear phase-specific capability boundaries. With precise context injection from the Verified Empirical Knowledge Base (VEKB), Hexstrike achieves a functional success rate (FSR) of 90.0%, indicating strong payload synthesis and iterative error recovery when reconnaissance noise is removed. In contrast, under unstructured reconnaissance without prior knowledge, targeted vulnerability recall (TVR) remains around 50%. When exposed to verbose scan outputs, agents exhibit attention drift and unreliable feature extraction, limiting their ability to localize vulnerable targets in noisy environments.

**Architectural trade-offs and vulnerability mapping.** State-management design determines the vulnerability classes each framework can handle effectively. Multi-agent systems (MAS), such as Hexstrike, enforce component isolation through the Model Context Protocol (MCP). By truncating exception stacks, they improve fault tolerance and are therefore well suited to node-crashing vulnerabilities, such as insecure deserialization. Monolithic architectures, such as Cyberstrike, avoid synchronization overhead and achieve high execution efficiency, but they are prone to context collapse in long interaction chains. This makes them more suitable for short-chain injection flaws that require only a small number of HTTP requests before context contamination occurs. Graph-driven frameworks, including RedAmon and Decepticon, maintain global state through structured multi-persona topologies. This design enables cross-session reasoning and comparison of privilege boundaries across accounts, making these systems effective for detecting business-logic flaws such as insecure direct object references (IDOR). Finally, constrained architectures, such as NeuroSploit, use symbolic attack trees and regex-based validation to reduce hallucinated tool invocations. However, this rigidity also limits payload mutation, confining their applicability to deterministic N-day verification based on standard proof-of-concept templates.

**Limitations and future work.** This study has two main limitations under the current technical setting. First, the benchmark relies on disclosed CVEs. While explicit knowledge injection improves reproduction of known vulnerabilities, it also highlights the models' dependence on prior exploit templates. To extend the decoupled paradigm to zero-day scenarios, future benchmarks could replace CVE-specific injection with abstract vulnerability semantics. When reconnaissance fails, the framework would inject only the vulnerable endpoint, parameter, and CWE category, rather than a known CVE identifier. This would require the agent to synthesize payloads without relying on historical proof-of-concept templates, enabling a more isolated assessment of zero-day exploit-generation capability. Second, the isolated sandbox does not model dynamic interaction with active defense mechanisms, such as endpoint detection and response (EDR) systems. Future agents could incorporate stronger planning methods, such as reinforcement learning, to improve evasion and long-horizon decision-making during high-risk adversarial interactions.

## 6 Conclusion

To mitigate error cascading in LLM-based penetration-testing evaluation, we propose a two-stage decoupled framework that integrates Ground-Truth Injection with a Verified Empirical Knowledge Base (VEKB). Our systematic benchmark clarifies the capability boundaries of LLM agents. With accurate contextual information, models demonstrate strong effectiveness in dynamic exploit execution, achieving an FSR of up to 90.0%. However, under unstructured reconnaissance, they remain vulnerable to attention drift and target-localization failures. Cross-architectural comparisons further show that state-management design induces distinct vulnerability niches across frameworks. Overall, the proposed decoupled paradigm offers an empirical basis for architecture selection and underscores a key transition in LLM-based offensive security: from static code generation toward autonomous exploit execution, where strict state isolation becomes essential.